# Control of Buckling of Colloidal Supraparticles

*Lukas J Roemling[1], Gaia De Angelis[2], Annika Mauch[1], Esther Amstad[2], Nicolas Vogel[1]\**

[1] Friedrich-Alexander-Universität Erlangen-Nürnberg, 91058 Erlangen, Germany

[2] Soft Materials Laboratory, École Polytechnique Fédérale de Lausanne (EPFL), 1015 Lausanne, Switzerland

**ABSTRACT.** Clusters of colloidal particles, often termed supraparticles, can provide more functionality than the individual particles they consist of. Since these functionalities are determined by the arrangement of the primary particles within a supraparticle, controlling the structure formation process is of fundamental importance. Here, we show how buckling is determined by particle-surfactant interactions and how the final morphology of the formed supraparticles can be controlled by manipulating these interactions in time. We use water/oil emulsions to template supraparticle formation and tailor the interactions of negatively charged colloidal particles with the surfactants stabilizing the water/oil-interface via the local pH within the aqueous droplet. At low pH, protonation of the anionic headgroup of the surfactant decreases electrostatic repulsion of the particles, facilitates interfacial adsorption, and subsequently causes buckling. We show that the local pH of the aqueous phase continuously changes during the assembly process. We gain control over the formation pathway by determining the point in time when interfacial adsorption is enabled, which we control via the initial pH. As a consequence, the final supraparticle morphology can be tailored at will, from fully buckled structures, via undulated surface morphologies to spherically rough and spherically smooth supraparticles and crystalline colloidal clusters.

**INTRODUCTION**

Colloidal particles are useful model systems to fundamentally study self-organisation phenomena[1–3] and provide functional materials with intriguing electronic[4–6], optical[7,8], or magnetic properties[9,10] due to their small size.

When such particles are arranged as colloidal crystals, entirely new properties emerge from the collective behaviour of those particles. For example, the constructive interference of light scattered at the constituent colloidal building blocks can result in structural coloration.[11–15] Supraparticles are interesting in this respect, as such emergent properties can be translated into well-defined, micron-scale building blocks that can be dispersed or used as a powder.[16–19] Supraparticles thus combine properties of individual building blocks and enable novel functionalities by colocalization, defined arrangement, or coupling between individual building blocks.[16,18,20] Since these emergent properties are directly related to the internal structure of such supraparticles, understanding and controlling the formation process is of general importance to reliably tailor supraparticle properties.

Supraparticles form by the self-assembly of colloidal primary particles during the drying of a particle-laden emulsion droplet. These droplets can be produced with different techniques. Spray drying is a fast and scalable technique that produces kinetically-controlled structures.[21] The drying velocity of droplets on superhydrophobic surfaces can be regulated by the humidity but is limited to comparably large supraparticles.[22,23] Emulsion templating allows the creation of much smaller droplets, and, with the aid of microfluidic devices, well-defined supraparticles with narrow size distributions can be produced.[24,25] In this system, the diffusion of water through the outer medium is required to consolidate the supraparticles[26–28] which enables very slow drying and the formation of highly crystalline clusters.[29,30]

Supraparticles from dried emulsion droplets can exhibit different structures and morphologies (Figure 1). Spherical supraparticles form when the primary particles are free to diffuse and consolidate within the confining droplet. By slowing down the self-assembly process, the resultant structure can resemble predicted minimum-energy structures[27,29–31] with high structural order and defined symmetry such as the icosahedral colloidal cluster shown in Figure 1a. Faster drying conditions typically produce spherical supraparticles with a well-ordered surface structure (Figure 1b).[32] Insufficiently stabilized primary particle dispersions often produce rough supraparticles without long-range order and random agglomeration of particles (Figure 1c).[33] However, a deviation from the spherical shape is also possible if buckling occurs during the drying process. We differentiate between rather spherical supraparticle with an undulated surface (Figure 1d) and the more extreme case of a fully collapsed, hollowed thin-sheets of colloidal particles, resembling a colloidosome (Figure 1e).[34]

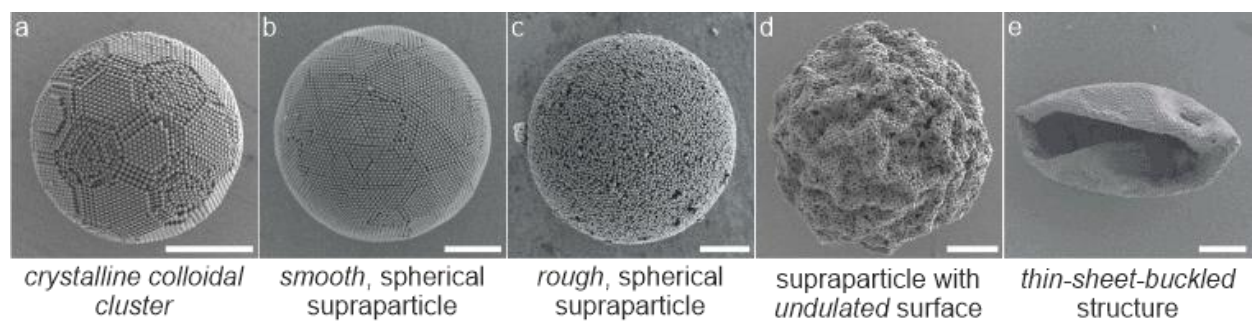

Figure 1. Supraparticles formed by the confined assembly of colloidal primary particles can show different morphologies. (a) Defined colloidal clusters with high degree of crystallinity and symmetry (a cluster with icosahedral symmetry is shown). (b) Smooth, spherical supraparticle with well-ordered surface structure. (c) Spherical supraparticle with a rough and disordered surface. (d) Supraparticle displaying an undulated surface. (e) Thin-sheet buckled supraparticle. All scale bars are 3 μm.

Buckling is associated with the ability of colloidal particles to adsorb to liquid interfaces. This adsorption is a function of the contact angle the particles assume at the interface[35] and thus relates to the relative affinity of the particles to both phases of an emulsion. The adhesion strength can vastly exceed kT, in effect causing irreversible particle adsorption.[35,36] This interfacial affinity can be exploited, *e.g.* for the design of Pickering emulsions[37–39], colloidosomes[34], particle-stabilized foams[40–44], or liquid marbles[45,46]. In these cases, the particles are often specifically designed[37] or surface-modified[47,48] to enable efficient interface adsorption. During the formation of supraparticles within emulsion droplets, any attraction of primary particles to the oil/water interface is likely to cause irreversible adsorption, producing a thin shell of particles and leading to the transition from viscous to elastic behaviour of the interface.[49] Such thin shells subsequently prevent the further consolidation of the colloidal primary particles during drying. Ultimately, the collapse of the thin shell thus results in dented or toroidal morphologies[50–53] or more generally in buckled structures.

In droplets produced in conventional microfluidic systems, the continuous phase often consists of fluorinated oils[27,31]. Most colloidal particles have a low affinity to fluorinated oils, making their adsorption at the drop interface energetically less favourable.[35] Nevertheless, buckled structures are still observed in such systems (Figure 1).[27] This observation lets us hypothesize that additional mechanisms must be involved to manipulate the interactions between primary particles and the interface of the emulsion. Here, we focus on the role of surfactants that are generally present in such emulsions to prevent droplet coalescence. However, their role in the self-assembly process of colloids has been largely overlooked. We elucidate how the particle-surfactant interactions control the affinity of the dispersed particles towards droplet interfaces. In particular, we exploit the pH responsive nature of Krytox FSH (Chemours), a typical fluorinated surfactant with a carboxylic acid end group that can be protonated at low pH values, to actively manipulate these interactions. Using charge-stabilized particles, we adjust the electrostatic repulsion of the particles via pH changes to control the particle affinity to the interface. Thereby, we exert control over the formation pathway of supraparticles and thus control their morphology.

**RESULTS AND DISCUSSION**

We form emulsion droplets of an aqueous dispersion of negatively-charged polystyrene (PS) particles in a continuous fluorinated oil phase by droplet-based microfluidics. As surfactants, we compare the anionic surfactant Krytox FSH, a perfluorinated polyether (PFPE) with a carboxylic acid head group, and a nominally non-ionic surfactant – a triblock of PFPE- O,O'-Bis(2-aminopropyl) polypropylene glycol-block-polyethylene glycol-block-polypropylene glycol (Jeffamine® ED-900)-PFPE (in the following abbreviated as PJP for PFPE-Jeffamine-PFPE) (Figure 2e). Both of these surfactants are commercially available and the PJP surfactant can be synthesized through an amide bond formation between Krytox FSH and Jeffamine to form the PJP block-copolymer.[54–56]

We hypothesize that the interaction of the surfactant with the dispersed particles determines the pathway of the supraparticle formation, as illustrated in Figure 2. For the combination of negatively-charged particles and negatively-charged Krytox FSH, we anticipate that the electrostatic repulsion effectively prevents interfacial adsorption, so that spherical, consolidated supraparticles can be formed (Figure 2a).

When the non-ionic PJP surfactant is used, no electrostatic interactions should be present to repel the particles. An affinity of the more hydrophobic polypropylene block to hydrophobic polystyrene moieties present at the PS particle surface is even suggested in literature.[57] We therefore hypothesize that the PS particles can adsorb to the liquid interface and thus produce buckled structures in the presence of PJP surfactants (Figure 2b). The same result is expected when we invert the particle charge and use positively-charged particles in combination with the PJP surfactant (Figure 2c). Finally, the combination of the negatively-charged Krytox FSH surfactant with positively-charged particles should result in a strong affinity that may bind the surfactant to the particle surface. We expect such Krytox FSH surface-modified particles to leave the aqueous droplet as individually dispersed particles into the fluorinated oil (Figure 2d).

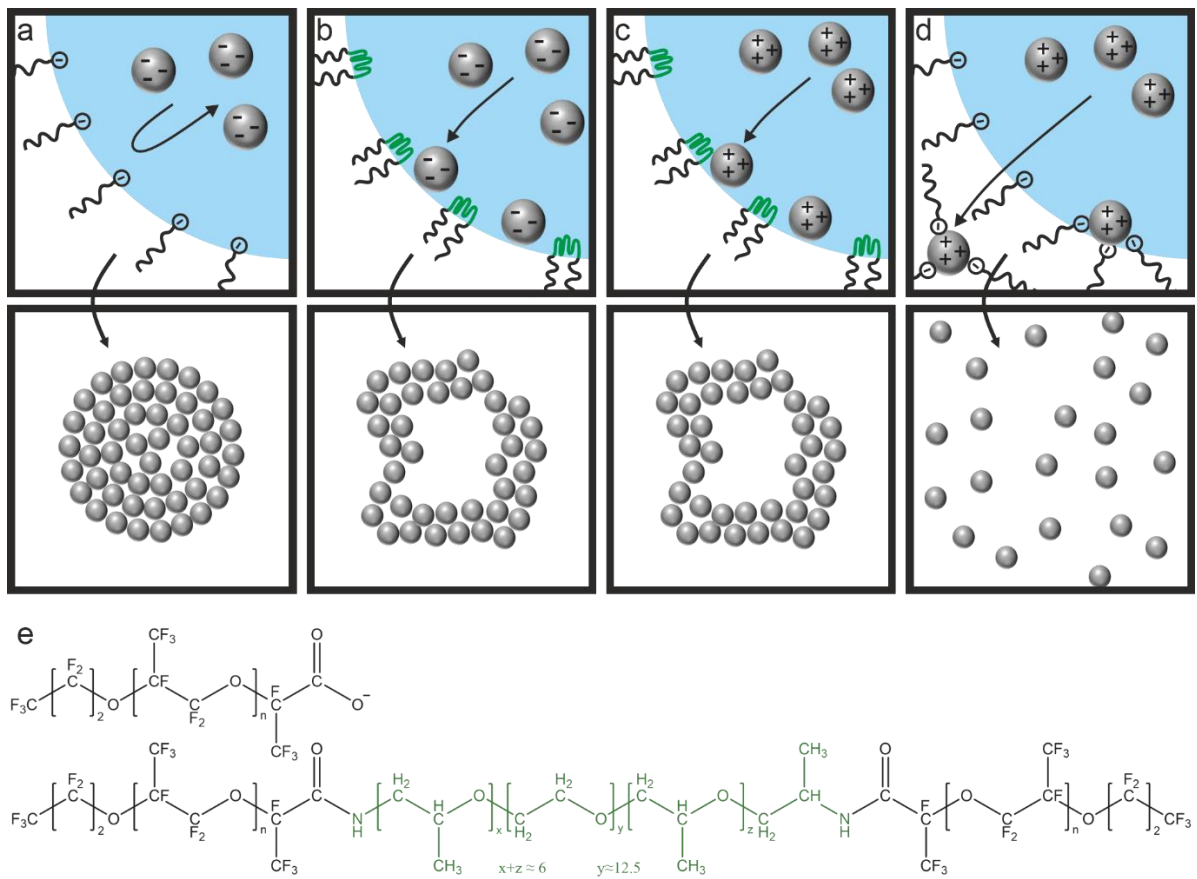

Figure 2. Hypothesis of supraparticle formation pathways as a function of surfactant/particle interactions. (a) Negatively charged particles in a droplet stabilized with an anionic surfactant should yield spherical supraparticles. (b,c) Non-ionic surfactant and either negatively (b) or positively (c) charged particles should produce buckled structures after the particles adsorb to the w/o interface. (d) Positively charged particles should be attracted by negatively-charged surfactant molecules, change their surface functionality and enable direct dispersion of the primary particles in the outer phase. The chemical structures of the two surfactants at neutral pH are shown in (e).

In all experiments of this study, we use 1 wt.-% of either positively or negatively charged particles and 0.1 wt.-% of the commercial anionic surfactant Krytox FSH or the commercial non-ionic PJP, shown in Figure 2e, unless denoted differently. The surfactant concentration was chosen to reliably produce stable emulsions but to avoid interference of the excess surfactant during characterization. Spherical supraparticles indeed form when using particles and surfactants with the same charge. The effective prevention of interfacial adsorption is evidenced by the consolidated, spherical particles as the one shown in Figure 1b (see Figure S1 for a statistical evaluation) and the well-defined colloidal cluster in Figure 1a (prepared with reduced droplet shrinkage rate). These results agree with different reports in literature on this system.[27,58,59]

Using oppositely-charged particles and surfactants results in the dispersion of primary particles in the outer fluorinated oil phase, as evidenced by a turbid continuous phase after a few minutes of mixing (Figure S2). This behaviour supports our hypothesis of an attractive surfactant/particle interaction (Figure 2d). To our surprise, however, emulsions formed with the non-ionic PJP surfactants did not show the expected buckling behaviour. Instead, the results resembled the samples prepared with Krytox FSH as anionic surfactant (Figure 3). Combinations of PJP with negatively-charged particles produced spherical supraparticles (Figure 3a), while positively-charged particles left the droplet and were dispersed as individual particles (Figure 3b).

This behaviour suggests that the non-ionic surfactant is impure and contains traces of anionic surfactant from the synthesis. We first measured the interfacial tension via the pendant drop method to investigate the ability of the two surfactants to adsorb to the interface (Figure S3). The decrease in interfacial tension from $31 \pm 1$ mN/m with 0.1 wt.-% Krytox FSH HFE/water to $18 \pm 1$ mN/m with 0.1 wt.-% PJP HFE/water demonstrates the superior efficiency of PJP in stabilizing the interface, caused by the increased size of the hydrophilic part of the block-copolymer.

However, a chemical analysis of the commercial PJP surfactant by Fourier-Transform Infrared (FTIR) spectroscopy (Figure 3c) shows the presence of a considerable amount of Krytox FSH residues, the starting material for the synthesis of the block copolymer.[60] The presence of free Krytox FSH in the commercial PJP block copolymer can be seen from the C=O stretch vibration of the carboxylic acid group. For pure Krytox FSH, this vibration occurs at ~1775 cm$^{-1}$ (Figure 3d, black curve). After amide coupling to Jeffamine to synthesize the non-ionic triblock, the energy of the amide-based C=O vibration shifts to ~1700 cm$^{-1}$.[61] This new peak is pronounced for a self-synthesized PJP surfactant (Figure 3c, dark green curve) we prepared following literature protocols.[54] In this case, only a minor carbonyl band associated with the carboxylic acid-based starting material is observed, indicating only small amounts of Krytox FSH impurities. For the commercial PJP, in contrast, the spectral signature of the amide bond only shows up as a shoulder, while the main peak occurs at the spectral position of the Krytox FSH starting material. Hence, we conclude that PJP surfactants can generally contain anionic components of Krytox FSH in varying quantities. This interpretation is further supported by $^{19}$F-NMR shown in Figure S4.[54] This chemical characterization rationalizes the behaviour of the dispersion droplets in Figure 3. The presence of anionic Krytox FSH facilitates consolidation into spherical supraparticles by repulsive interaction with negatively-charged particles (Figure 3a) but cannot confine positively-charged particles within the droplet (Figure 3b). Note that the behaviour of the synthesized surfactant with presumably higher purity was similar in the same emulsion system (Figure S5).

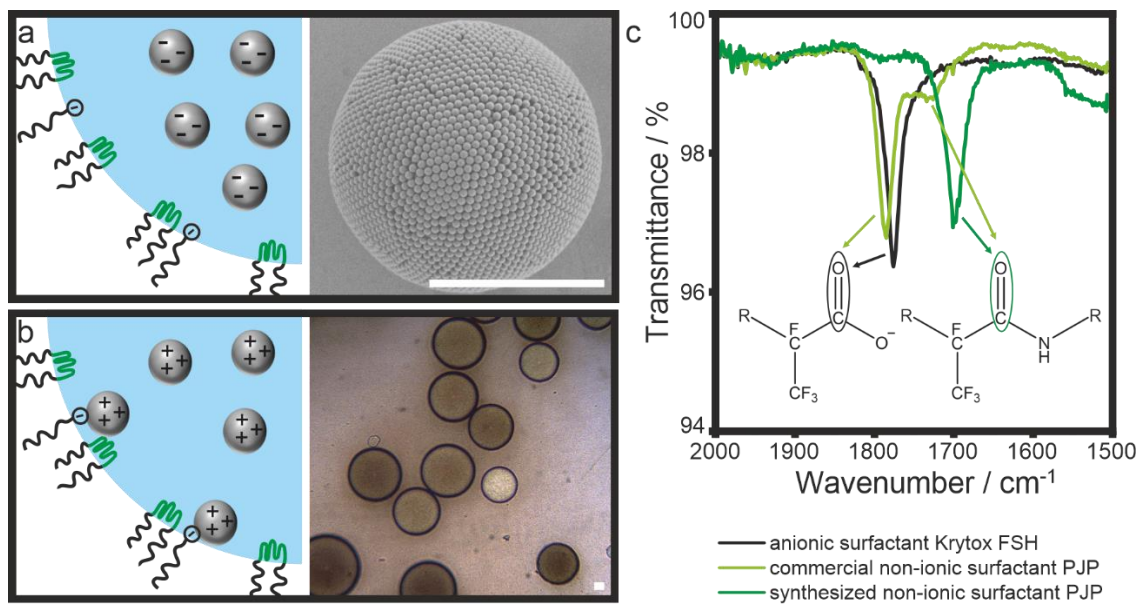

Figure 3. Supraparticles formed within emulsion drops stabilized with the non-ionic commercial surfactant PJP and characterization of the different surfactants. (a) The combination of PJP with negatively-charged particles produces spherical supraparticles. (b) The combination of PJP with positively-charged particles leads to individual primary particles leaving the droplet and to become dispersed in the continuous phase. (c) FTIR spectra of the three surfactants, showing the carbonyl absorption band associated with carboxylic acids, and amides, respectively. All scale bars are 5 µm.

To investigate the behaviour of the pure non-ionic surfactant, and, in particular, to probe if the reduced repulsion causes the formation of buckled supraparticles, requires a different approach. We capitalize on the possibility to affect the surfactant properties of Krytox FSH using the pH of the aqueous phase. If the pH of the inner water phase is below the $pK_A$-value of the carboxylic acid headgroup, protonation of the group will eliminate the anionic charge, thus reducing its impact on the structure formation process.

The pK$_A$-values for carboxylic acids in different chemical environments are reported to range from 2 to 4.[62] The interfacial tension between a fluorinated oil droplet containing 0.1 wt.-% Krytox FSH and water shows a significant increase if the pH of the aqueous phase is shifted from 5 to 3 (Figure S6), indicative of the reduced hydrophilicity of the headgroup. This trend is not observed for the commercial PJP surfactant, indicating that in this case, the non-ionic block-copolymer remains interfacially active at all pH values (Figure S6). As even small amounts of surfactant impurities can significantly change the assembly process (Figure 3a,b), we anticipate that adjusting the pH will enable us to control the surfactant/particle interactions in our system by manipulating electrostatic repulsion.

Additionally, due to the nonpolar character of the continuous fluorinated oil phase, we do not expect charged molecules and particles to be able to leave the aqueous phase. We therefore hypothesize that the diffusion of water molecules into the oil phase during droplet shrinkage continuously changes the pH within the water droplet of the emulsion. Thus, when starting with an initial moderate concentration of H$^+$ ions (*i.e.* a pH above the pK$_A$ value of the carboxylic acid headgroup), the proton concentration will continuously increase, and the pH value correspondently decrease as water molecules diffuse out of the droplet during the drying process. We test this hypothesis using emulsions of pure water droplets with the pH indicator bromophenol blue in a continuous fluorinated oil phase (Figure 4). Bromophenol blue shows a colour change from purple to yellow between pH 4 and pH 2 (Figure 4a). We adjust the initial conditions to pH = 4 and the formed emulsions show a purple colour (Figure 4b). In the course of the drying process, the colour of the emulsion changes to pale yellow (Figure 4b), demonstrating that the local pH within the emulsion droplets indeed changes upon droplet shrinkage.

This ability to manipulate the local pH provides us with a handle to control the point at which the surfactant properties in the spherical confinement change, as shown in Figure 4c. Thereby, we can manipulate the confined self-assembly process at will. We expect that at moderate or high pH values (pH>$pK_A$ (Krytox FSH)), negatively charged particles will be repelled from the interface due to the electrostatic repulsion of the deprotonated carboxylic acid headgroup of the surfactant. At low pH we expect the particles to adhere to the w/o-interface due to a lack of repulsive forces and their affinity to the PEG-PPG-PEG moiety of the non-ionic surfactant.[57] To support this idea, we use interfacial rheology to investigate the affinity of the charge-stabilized particles to the surfactant-covered interface at two different pH values that are far below the $pK_A$ (pH = 1.3) and far above the $pK_A$ (pH = 5) (Figure 4d). At pH 1.3, where nearly all surfactant molecules are uncharged, the 2D storage (G') and the loss moduli (G'') increase after 90 minutes, indicating and increased elasticity of the interface due to the adsorption of particles.[63] At pH = 5, where any Krytox FSH residues are negatively charged, the storage and loss moduli remain constant for the entire experimental time frame of 5 hours, indicating that the nature of the interface did not change. With charges present in the surfactant head group (pH = 5), we therefore conclude that interfacial adsorption is prevented. Conclusively, if the pH is chosen to be below the $pK_A$ value of the surfactant, protonation of the carboxylic acid head group removes charges and increases the interfacial affinity of the particles. As a result, a monolayer of particles adsorbs to the oil/water interface.

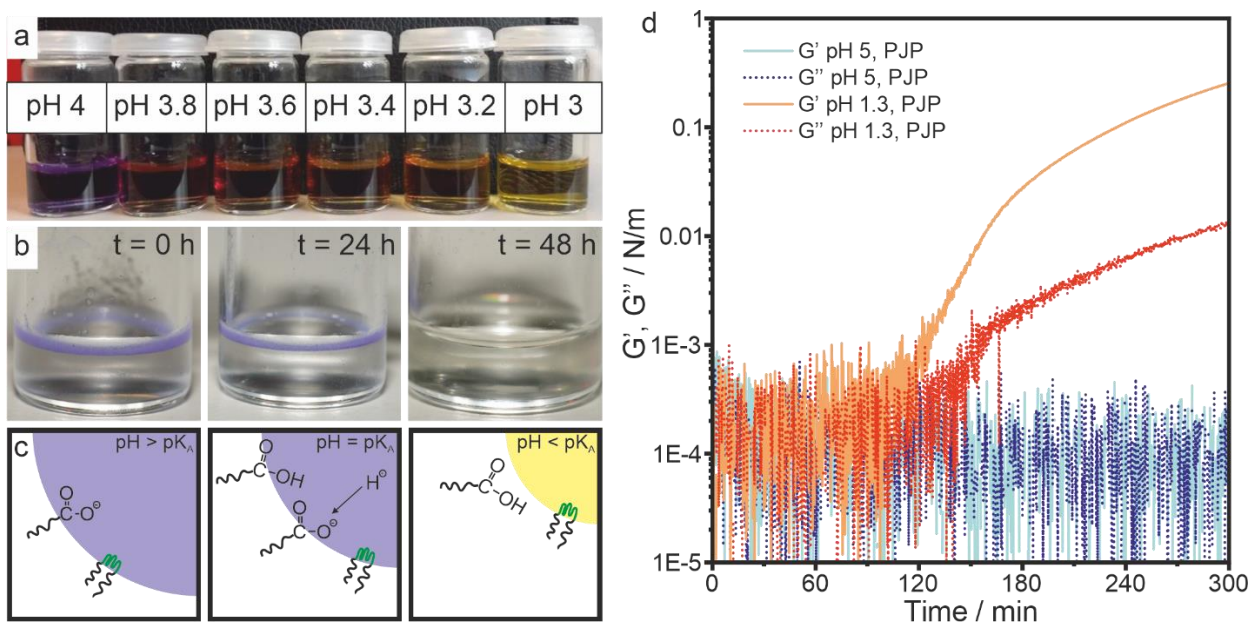

Figure 4. The local pH changes within emulsion droplets upon shrinkage in water/fluorinated oil emulsions, affecting the affinity of particles to the w/o interface. a) Aqueous solutions of bromophenol blue show a colour change from violet/blue to yellow between pH = 4 and pH = 3. b) Evolution of the pH inside w/o emulsion droplets over time, shown by the colour change of bromophenol blue. c) Schematic illustration of the protonation of the PJP surfactant at the water-oil interface during drying. (d) Interfacial rheology shows that the storage (solid lines (orange/light blue)) and loss moduli (dotted lines (red/dark blue)) of an interface between an aqueous particle dispersion and the fluorinated oil phase, stabilized by 0.1 wt.- PJP (commercial) increase at pH 1.3 (below the p$K_A$ of Krytox FSH) but are constant at pH 5 (above the p$K_A$ of Krytox FSH), indicating particle adhesion to the interface at low pH.

We established that we can control the pH inside emulsion droplets upon shrinkage, the surfactant charge via the pH, and thus the adsorption of colloidal particles at the oil/water interface. By combining these properties, we control the formation pathway, and thus the final morphology of supraparticles. We adjust the initial pH within droplets of an aqueous colloidal dispersion (0.1 wt.-% negatively charged PS particles) by centrifugation of the primary particles and their redispersion in an aqueous solution with a fixed pH between 2 and 4, adjusted by the addition of HCl. We use droplet-based microfluidics[18] to produce uniform emulsion droplets within a continuous fluorinated oil phase containing the commercial PJP surfactant with a concentration of 0.1 wt.-%. The droplets shrink at room temperature for 24 hours and we characterize the resulting morphologies using statistical evaluation of Scanning Electron Microscopy (SEM) images (Figure 5). For the evaluation we distinguish four classes of supraparticle morphologies, as introduced in Figure 1b-e: *smooth* supraparticles with a well-defined spherical shape and ordered surface structure; *rough* supraparticles with a consolidated spherical shape but no visible ordered at their surface; *undulated* supraparticles with a near-spherical shape but pronounced buckled features at the surface; and *thin-sheet buckled* supraparticles with a surface that is reminiscent of a thin sheet that crumbled under pressure.

As shown in Figure 5, an initial pH value of 2 to 2.25 resulted in the almost exclusive appearance of thin-sheet buckled supraparticles. Increasing the initial pH value to 2.5 predominantly produced undulated supraparticles. A further increase in pH to 2.75 resulted in mostly rough supraparticles. At an initial pH = 3, only spherical supraparticles were observed. Their topology is either rough or smooth, with both morphologies occurring in close to equal ratios. Finally, with an initial pH of 4 smooth supraparticles exclusively formed. Note that such supraparticles can be further crystallized into well-ordered colloidal clusters with defined geometries, such as the icosahedral cluster shown in Figure 1a.[27,29]

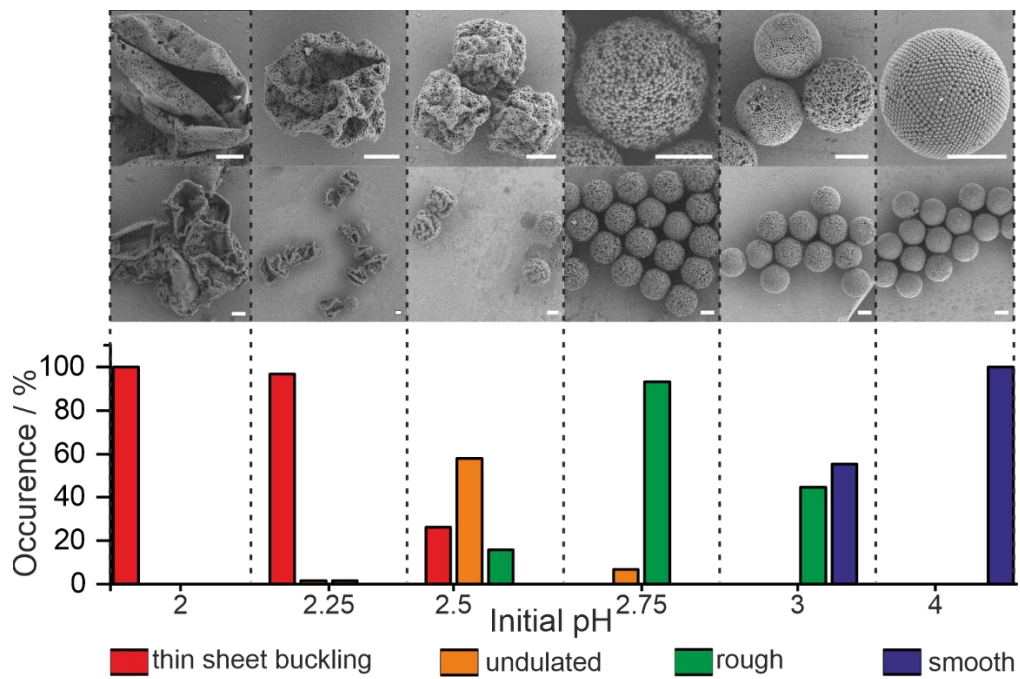

Figure 5. Control of supraparticle morphology via the initial pH within the emulsion droplet, characterized via statistical analysis of SEM images. All scale bars are 3 μm.

The resulting supraparticle morphologies are deterministically controlled via the initial pH because it controls the point in time at which the primary particles are able to adsorb to the droplet interface, as schematically illustrated in Figure 6b. If the initial pH value is only slightly above the pK$_A$ value of the carboxylic acid group of the surfactant, protonation occurs early in the formation process, and most particles will adsorb to the w/o interface resulting in a thin sheet buckled supraparticle (Figure 6b, top row). This is the case for initial pH < 2.25, as shown in Figure 5. If the initial pH value is higher than the pK$_A$ value of the respective carboxylic acid group (Figure 6b, bottom row), e.g. for an initial pH = 2.5, the particles will start to adsorb to the interface at a later stage of the formation process, when the shrinking droplet reaches the pK$_A$ (Figure 6b). As a result, buckling will occur when the size of the droplet is already reduced, and the available interfacial area is smaller. Thus, less particles can adsorb and the droplet does not reduce much more in size, leading to undulated supraparticle morphologies after drying (Figure 6b). Finally, when the pH never decreases below the pK$_A$, spherical supraparticles result as the primary particles can consolidate within the spherical confinement without any interference from the liquid interface (Figure 6c), as is the case for an initial pH ≥ 4.

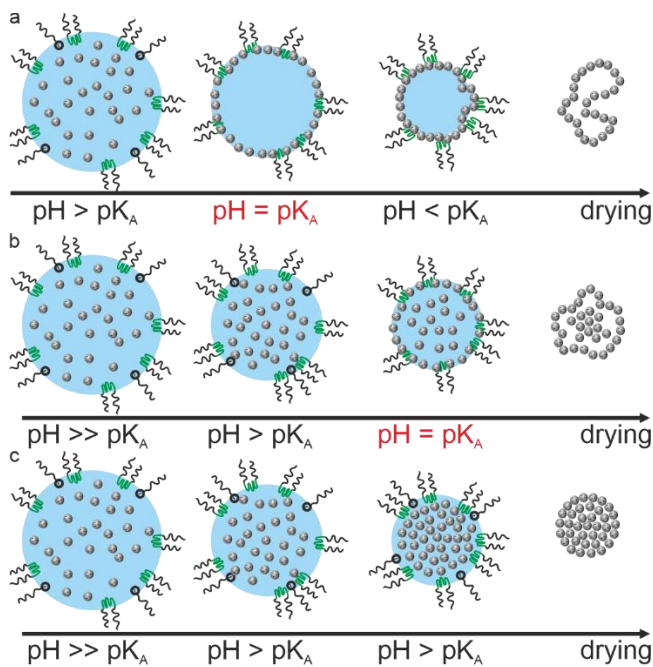

Figure 6. Schematic representation of the drying process of a w/o emulsion droplet containing colloidal particles with different starting pH values. (a) At a low initial pH, the pK$_A$ is reached early in the supraparticle formation process and particles adsorb to the interface. Further drying leads to thin-sheet buckled structures. (b) A moderate initial pH leads to adhesion of the particles at a later stage of the formation process. Since less interface is available, more particles remain in the bulk, and the droplet does not shrink much more before consolidation, a more spherical shape is retained and an undulated morphology is observed. (c) A high initial pH prevents the adhesion of the particles during the entire formation process and fully spherical supraparticles are formed.

Having established the formation mechanism, we can leverage other experimental parameters to control supraparticle morphologies. We tune the final morphology with the droplet size. Larger droplets have a lower surface-to-volume ratio than smaller droplets. With conditions favouring particle adsorption to the interface early in the formation process (protonated Krytox FSH surfactant; pH=2.5), the interface will be readily covered with particles, templating buckled morphologies. With increasing droplet size, the surface-to-volume ratio decreases, leaving a higher ratio of particles in the bulk when the interface is saturated with particles. The degree of buckling should thus be less pronounced (Figure S7a). We demonstrate this control by adjusting the initial droplet size using droplet-based microfluidics, using a fixed primary particle concentration and statistically evaluate the supraparticle morphologies via SEM image analysis as a function of the initial emulsion droplet size (Figure 7a). We observe a shift from thin-sheet buckled supraparticles as the main morphology for small initial droplet sizes (20 μm), via predominantly undulated morphologies (25 μm - 31 μm initial droplet sizes) to rough supraparticles (46 μm initial droplet size). In a similar fashion, the resultant morphology depends on the initial primary particle concentration, since it determines the number of particles that remain in the bulk liquid after interfacial adsorption. When increasing the particle concentration from 1 wt.-% to 5 wt.-% with similar initial droplet sizes (20 μm) and constant surfactant concentration (0.1 wt.-%), the main fraction of observed morphologies shifts from thin sheet buckled supraparticles, to undulated and rough supraparticles (Figure 7b, Figure S7). While with the low particle concentration, nearly all primary particles adsorb to the liquid interface, and thus produce a fully buckled morphology, the increased number of primary particles retains particles in the bulk and thus decreases the degree of buckling upon consolidation.

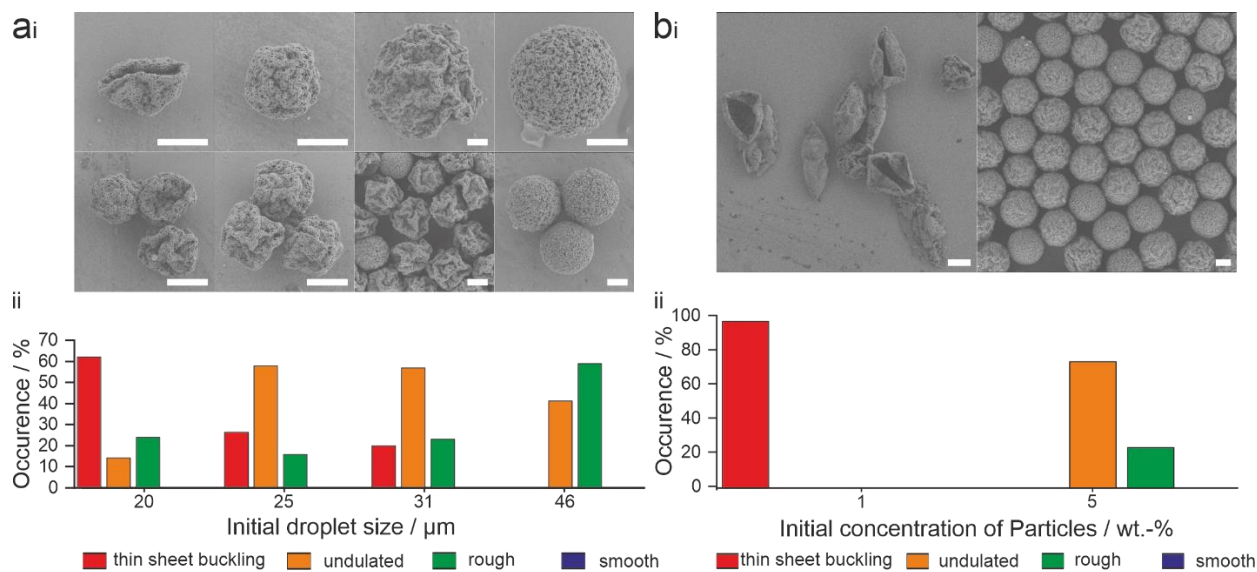

Figure 7. Control of supraparticle morphology via the number of primary particles available in the bulk of the droplet. (a) Influence of droplet size. W/O emulsions were produced at an initial pH = 2.5 with different droplet sizes at the same particle concentration (1 wt-%). (i) representative SEM images of the formed supraparticles. (ii) statistical evaluation of the occurrence of characteristic morphologies. (b) Influence of solid content. W/O emulsions were prepared with an initial droplet size of 20 μm, at an initial pH = 2 with different primary particle concentrations. (i) representative SEM images of the formed supraparticles. (ii) statistical evaluation of the occurrence of characteristic morphologies. All scale bars are 5 μm.

**CONCLUSION**

Supraparticles formed by the confined assembly of colloidal primary particles within emulsion droplets can exhibit different morphologies. We identify the particle/surfactant interactions as the determining factor to control these morphologies. Efficient repulsion, caused by like-charged systems produce spherical supraparticles. Insufficient repulsion or weak attraction causes interfacial adsorption of the particles, triggering buckling. Strong attraction, for example using oppositely-charged systems do not produce stable supraparticles as surfactant-coated primary particles can leave the droplet.

Two key parameters can be generalized to successfully predict or understand buckling in colloidal supraparticle formation. First, if particles do not adhere to the interface, no buckling will be observed. If particles adhere to the interface, the irreversible adsorption causes buckling upon further shrinkage. Second, the point in time at which the particles adsorb to the interface during the formation process and how many particles are left in the bulk of the droplet will determine the resulting morphology. In well-established water-in-fluorinated oil emulsions stabilized with Krytox-based surfactants, we found that non-ionic fluorinated surfactants contained traces of anionic surfactant from the synthesis. The point in the formation process at which the interfacial adsorption occurs can be controlled via the protonation of the charged surfactant headgroup, triggered by a continuous change in pH within the emulsion droplet. The number of particles left in the bulk can be controlled by adjusting the particle concentration in the droplet, or by changing the surface-to-volume ratio via the droplet size. These insights provide simple handles to reliably control the morphology and thus the properties of supraparticles formed by emulsion templating.

## MATERIALS AND METHODS

**Synthesis of charge-stabilized polystyrene colloidal particles.** The synthesis was adopted from literature.[64] In brief, 250 mL of water was heated up to a temperature of 80 °C in a three-necked-flask equipped with a reflux condenser. The system was constantly flushed with nitrogen. After heating, 10 g of styrene (> 99 %, Sigma Aldrich) was added. After another 10 min, 0.052 g of sodium 4-vinylbenzenesulfonate (> 90 %, Sigma Aldrich), dissolved in 5 mL of water was added as a comonomer and stirred for 5 minutes before 0.1 g of ammonium persulfate (> 98 %, Sigma Aldrich) dissolved in 5 mL of water was added. After 22 h, the nitrogen flow and heating were stopped, and the colloid was left to cool. The resulting polymer colloidal dispersion was filtered through a lint-free tissue, extensively dialyzed and resulted in particles of the size d=246±2 nm and a zeta potential of $\zeta$=- 35 mV. Alternatively, positively charged particles were synthesized by using (Vinylbenzyl)trimethylammonium chloride (>99%, Sigma Aldrich) as comonomer instead. The resulting particles showed a size of d=288±4 nm and a zeta potential of $\zeta$=+31 mV.

**Synthesis of non-ionic fluorinated surfactant.** The non-ionic triblock surfactant of Krytox FSH - Jeffamine900 - Krytox FSH was bought from Creative PEGWorks and used as received and compared against the same surfactant synthesized following a protocol from literature.[54] For this, 5 g of Krytox FSH (Chemours) were stirred under Argon (Ar) atmosphere. One drop of anhydrous DMF (>99%, Sigma Aldrich) and 0.8 mL of 2M oxalyl chloride in DCM (Sigma Aldrich) were added and the mixture was stirred for 4 hours. Afterwards, the DCM and leftover oxalyl chloride were removed by rotary evaporation (Hei-VAP, Heidolph, Germany) for 2 hours, at 40 degrees and 1 mbar. The product was dissolved in 10 mL Novec™ 7100 (3M) and set under Ar atmosphere and 0.5 molar equivalent of dry Jeffamine900 (Sigma Aldrich) dissolved in 5 mL anhydrous DCM (> 99.8 %, Sigma Aldrich) was added. The reaction was left to reflux at 65 °C overnight. To purify the surfactant 2 mL of Novec™ 7100 (3M) and 50 mL of Methanol (> 99 %, Sigma Aldrich) were added to precipitate the block copolymer, which was centrifuged at 3 °C and 3000 RPM (Mega Star, 1.6R, VWR). The supernatant was discarded and the cleaning procedure was repeated four times to afford a pale white product.

**Production of Microfluidic channels.** Sylgard 184™ was purchased from Biesterfeld AG. Resin and crosslinker were mixed in a ratio of 10:1, degassed for 10 min under vacuum, and subsequently poured on a Silicon master. It was cured at 80 °C for 24 h. Afterwards, the PDMS was removed from the mold, activated in $O_2$-plasma for 30 s, and bonded on an activated glass slide. The channels were then flushed with 3 wt.-% Trichloro(1H,1H,2H,2H-perfluorooctyl)silane (Sigma Aldrich, 97%) in Novec™ HFE 7500 (3M) to turn the channels fluorophilic.

**Preparation of glass vials.** 1.5 mL glass vials (Carl Roth) were placed in a beaker filled with denatured ethanol (Carl Roth, 96%) and ultrasonicated for 1 min. The glass vials were then dried and activated in $O_2$-plasma for 5 minutes. Afterwards the vials are placed in a desiccator and 50 μL of Trichloro(1H,1H,2H,2H-perfluorooctyl)silane was added to the chamber. After storage under vacuum for 24 h the vials were heated to 80 °C for 2 hours and rinsed with denatured ethanol to get rid of excess silane.

**Supraparticle formation.** The microfluidic channel consists of 2 inlets and 1 outlet. The inlet for the inner phase is connected to a syringe pump with the aqueous particle-laden phase, while the inlet for the outer phase is connected to the outer fluorinated phase, which consists of 0.1 wt.-% PFPE5000-Jeffamine900-PFPE5000 (Creative PEGWorks) in Novec™ HFE 7500 (3M). The flow velocity could be adjusted separately. The resulting droplet size was adjusted by using two different channel sizes (15 μm or 25 μm) and varying the flow rates of the water/oil phase between 50/800, 50/200, and 200/100 μL/h. The resulting emulsion was collected in a pipette tip and subsequently transferred to a sample vial with a fluorinated surface to avoid droplet coalescence with the glass surface. The vial was stored without caps at room temperature to allow water to diffuse out of the droplets and evaporate to induce assembly. The formation process took approximately 24-36 hours.

**Characterization.** The supraparticles were characterized using an GeminiSEM 500 (Carl Zeiss, Germany) and the software ImageJ. Surfactants were characterized using FTIR (Spectrum 3, PerkinElmer), NMR spectroscopy (Bruker AVANCETM-400 spectrometer operating at 400 MHz. Samples are dissolved at a concentration of 10 mg in 0.75 mL of hexafluorobenzene before analysis), pendant drop measurements (DSA 30 drop shape analyzer, Krüss) and interfacial rheology (DHR-3, TA Instruments) using a double – wall ring (DWR) geometry.

## ASSOCIATED CONTENT

The following files are available free of charge.

Supporting information (PDF)

## AUTHOR INFORMATION

### Corresponding Author

\* nicolas.vogel@fau.de

### Author Contributions

The project was conceptualized by LJR and NV. A selection of suitable surfactants was done by LR, GDA, NV and EA. AM produced pH dilutions and screened the evolution of pH in the emulsion droplet. GDA synthesized the fluorinated surfactant PJP and performed IR and NMR measurements. The results were discussed between all authors. LR wrote the manuscript with input from all authors. All authors have given their approval to the final version of the manuscript.

**Funding Sources**

This research was funded by the Deutsche Forschungsgemeinschaft (DFG, German Research Foundation) – Project-ID 416229255 – SFB 1411

**ACKNOWLEDGMENT**

The authors would also like to thank Aurélien Bornet for his help with NMR analysis as well as Natalie Bonakdar and Gudrun Bleyer for supplying the negatively and positively charged PS particles respectively.

# Supporting Information:

# Control of Buckling of Colloidal Supraparticles

*Lukas J Roemling[1], Gaia De Angelis[2], Annika Mauch[1], Esther Amstad[2], Nicolas Vogel[1]\**

[1] Friedrich-Alexander-Universität Erlangen-Nürnberg, 91058 Erlangen, Germany

[2] École Polytechnique Fédérale de Lausanne (EPFL),1015 Lausanne, Switzerland

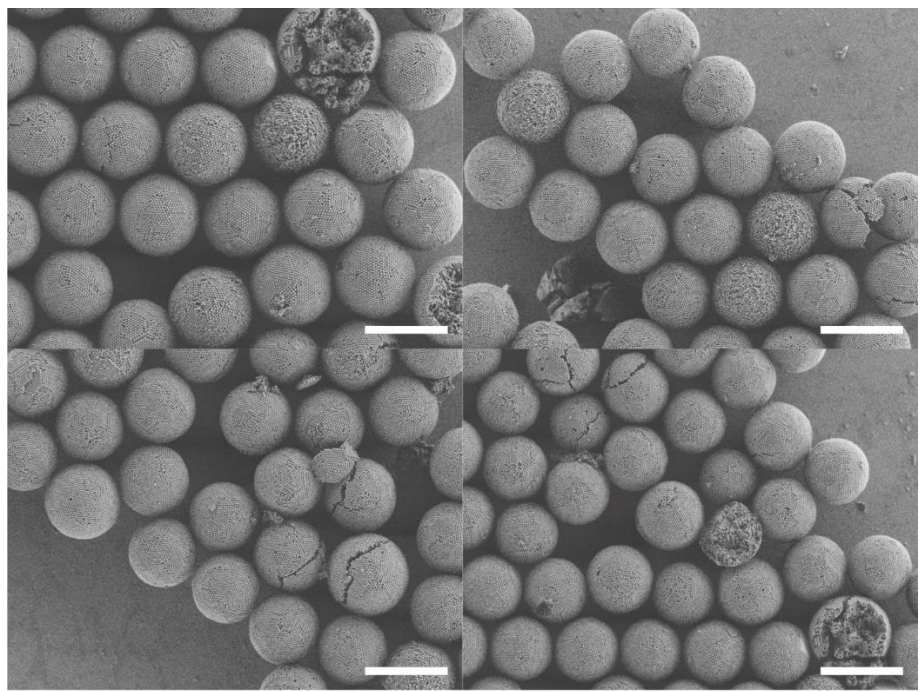

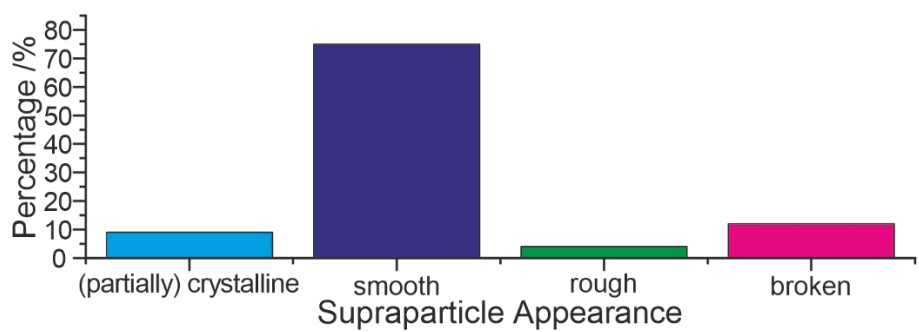

Figure S 1. Low-resolution SEM images of supraparticles produced via microfluidics using negatively charged polystyrene primary particles and anionic Krytox FSH surfactant. 75% of the cluster appeared to have a smooth surface, 9% were partially crystalline clusters, 12 % of the supraparticles were broken, probably due to the deposition (after the consolidation of the supraparticle). Only 4% of the supraparticles showed a rough surface without apparent order. Scale bars are 10 μm.

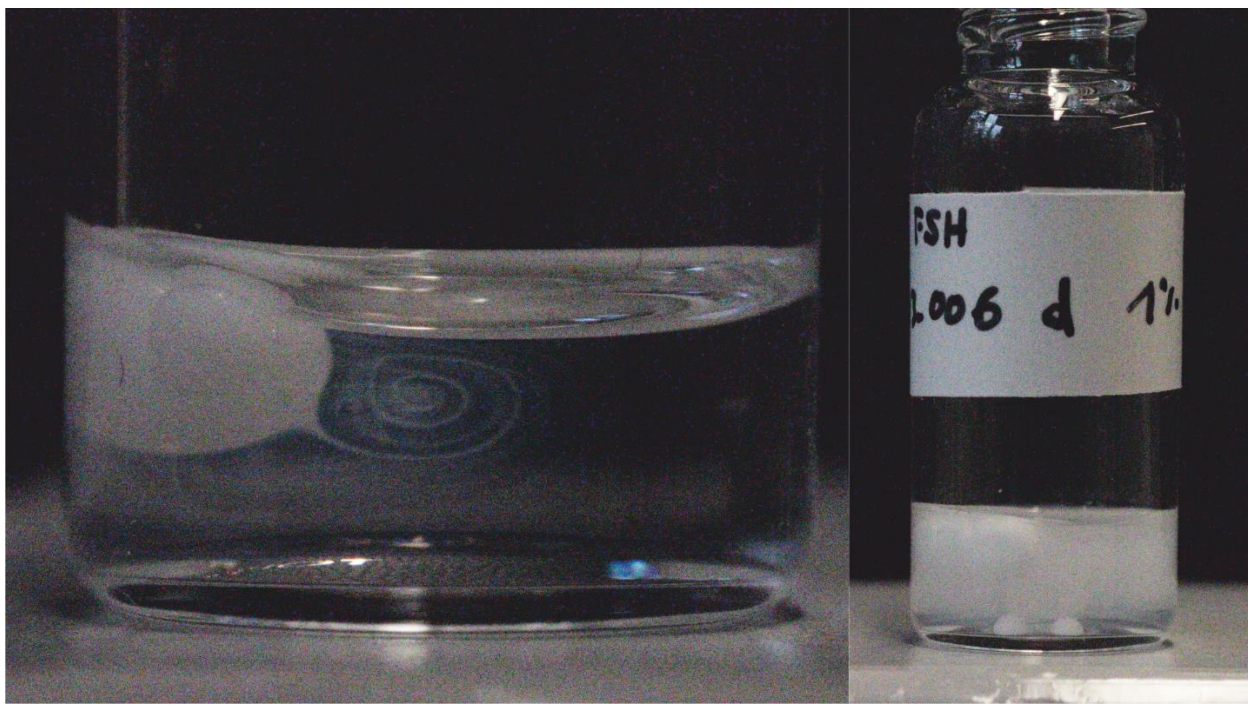

Figure S 2. Photographs of water-in-oil emulsion with positively charged particles in the water droplet. The oil phase contains anionic Krytox FSH. Shortly after emulsification the continuous outer oil phase becomes turbid due to particles dispersing in the fluorinated oil. After several minutes, the entire oil phase is opaque.

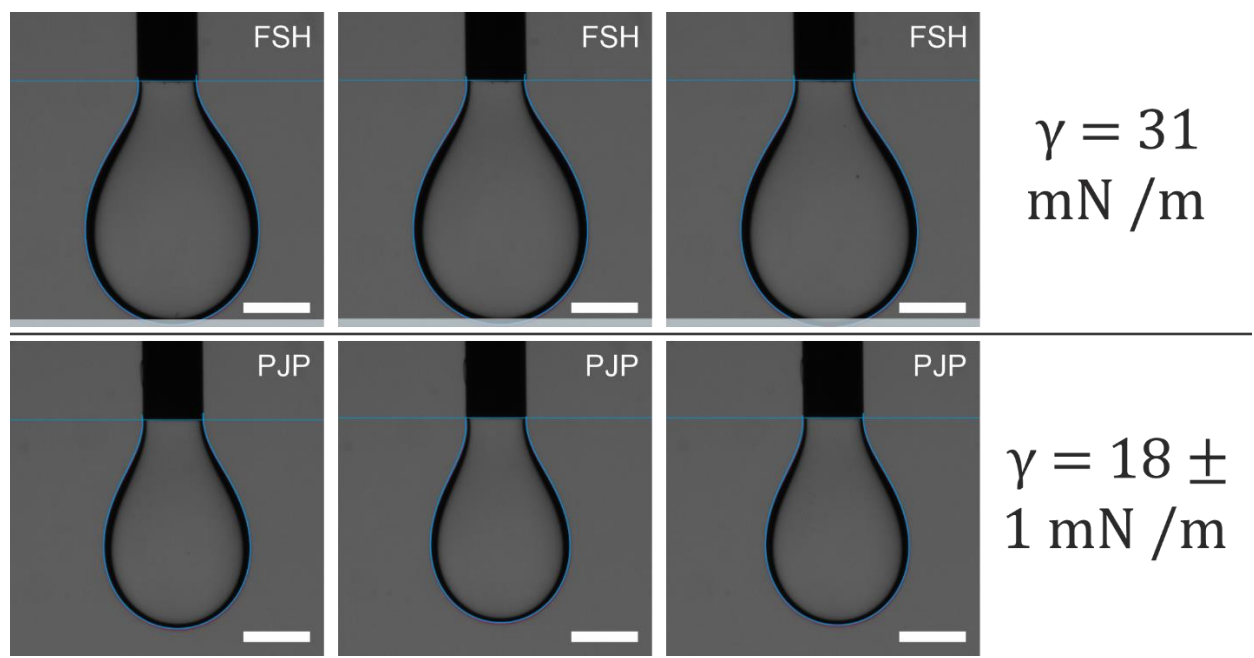

Figure S 3. Surface tensions of droplets of fluorinated oil (HFE) in water with 0.1 wt.-% of different surfactants (top: Krytox FSH; bottom: PJP triblock-copolymer) measured using the pendant drop method. PJP reduces the interfacial tension between the phases considerably more than FSH. Scale bars are 1 mm.

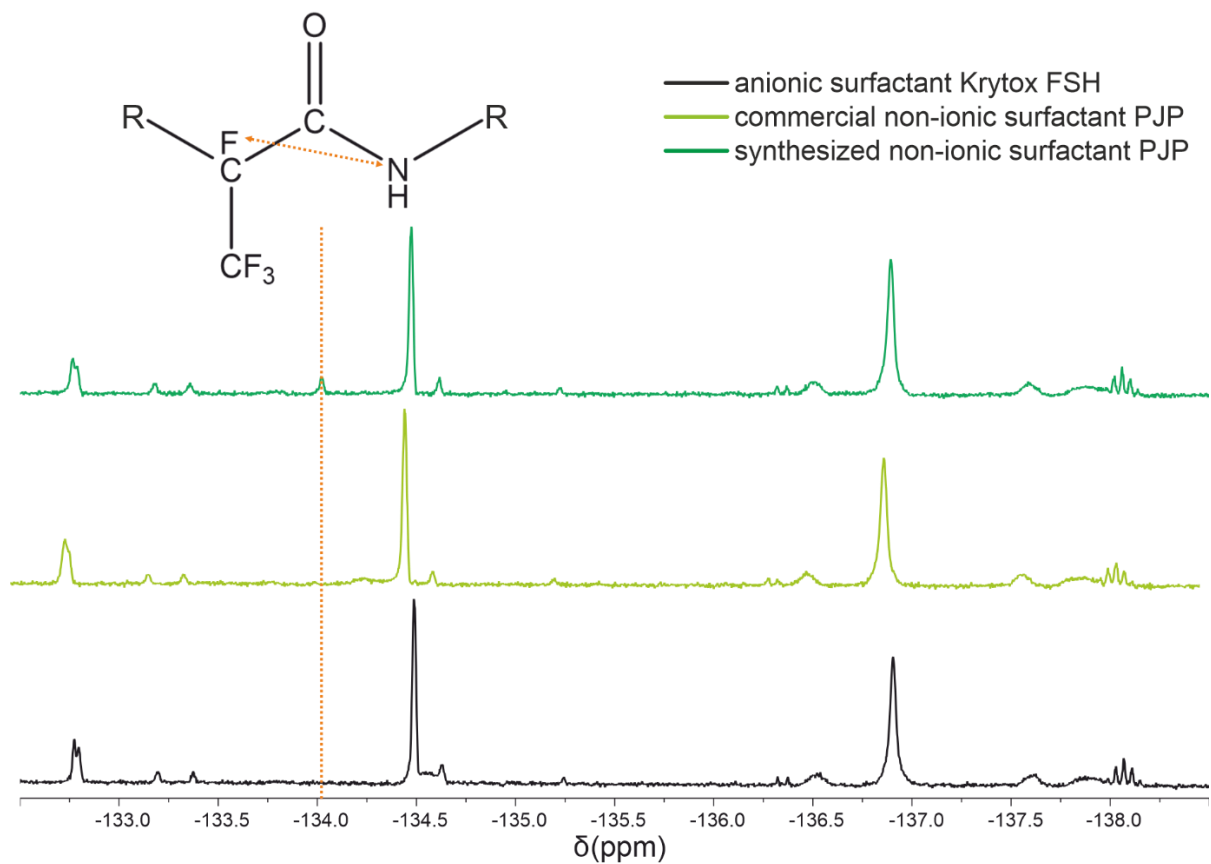

Figure S 4. NMR spectra of the three different surfactants used in this study. A signal at -134.025 ppm can be observed for the synthesized PJP surfactant. We attribute this peak to coupling between the F-atom at the α-C and the N-Atom, indicated by the orange arrow. The absence of this peak in the commercially available surfactant supports our hypothesis of large amounts of anionic Krytox FSH impurities, corroborating the FT-IR investigation shown in Figure 4.

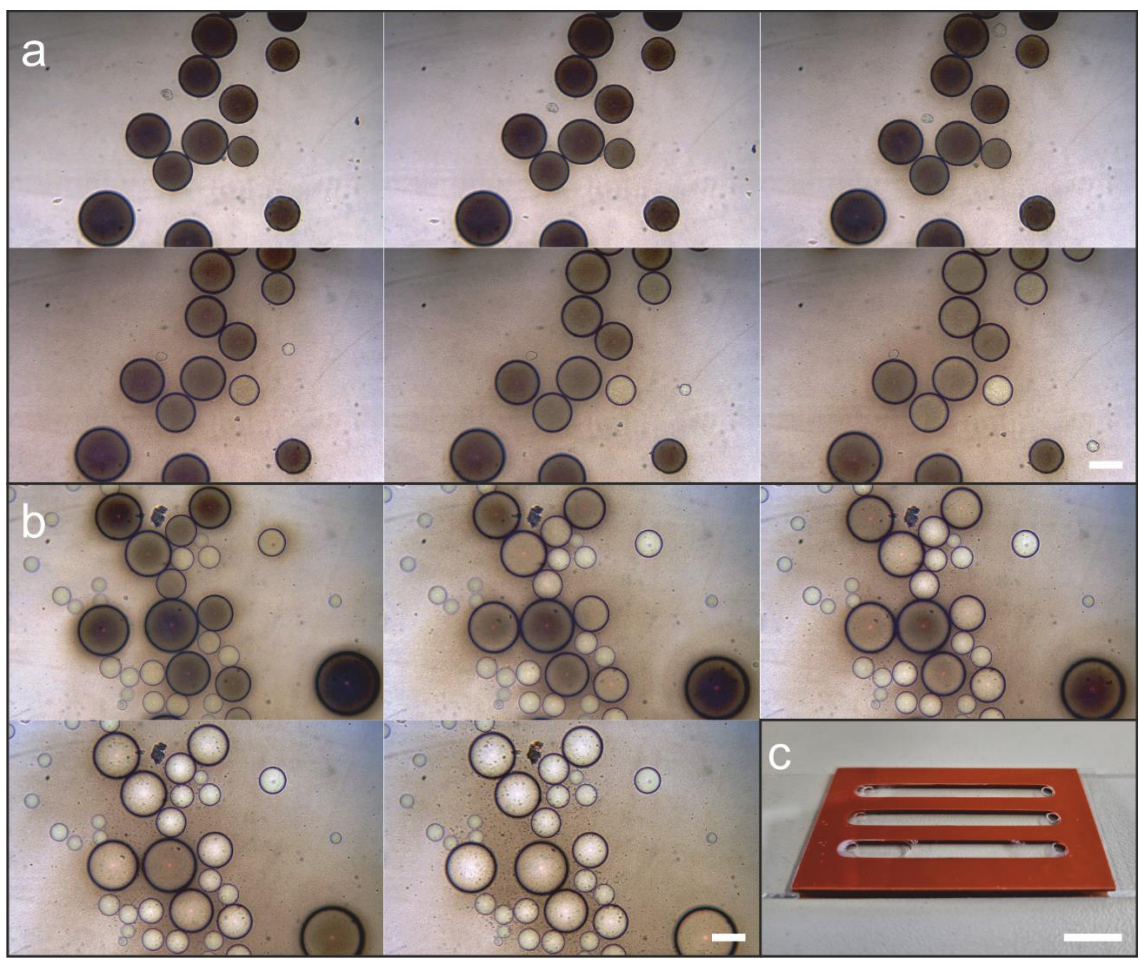

Figure S 5. Light microscopy images of aqueous droplets laden with positively charged particles in fluorinated oil stabilized by non-ionic surfactants. The emulsions were stabilized with (a) 0.1 wt.-% commercial PJP or (b) 0.1 wt.-% self-synthesized PJP. In both cases particles, the microscopy images show particles leaving the droplets over time (left to right, top to bottom). (c) The emulsions were produced by vortexing and then transferred to a capillary shown here to avoid fast drying in air. Scale bars are 20 μm (a+b) and 10 mm (c).

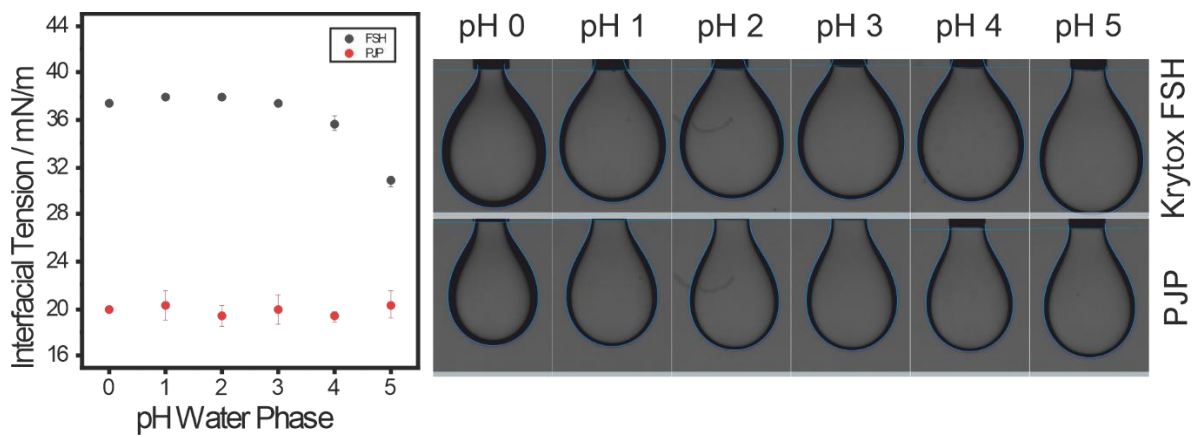

Figure S 6. Interfacial tension between a droplet of fluorinated oil (Novec HFE 7500) containing 0.1 wt.-% of either Krytox FSH or PJP surfactant and water at different pH values. The interfacial tension was generally lower for the PJP surfactant and remained constant at all pH values. The anionic Krytox stabilizes the emulsion well at moderate pH values but is less interfacially active at low pH due to the protonation of the carboxylic acid headgroup.

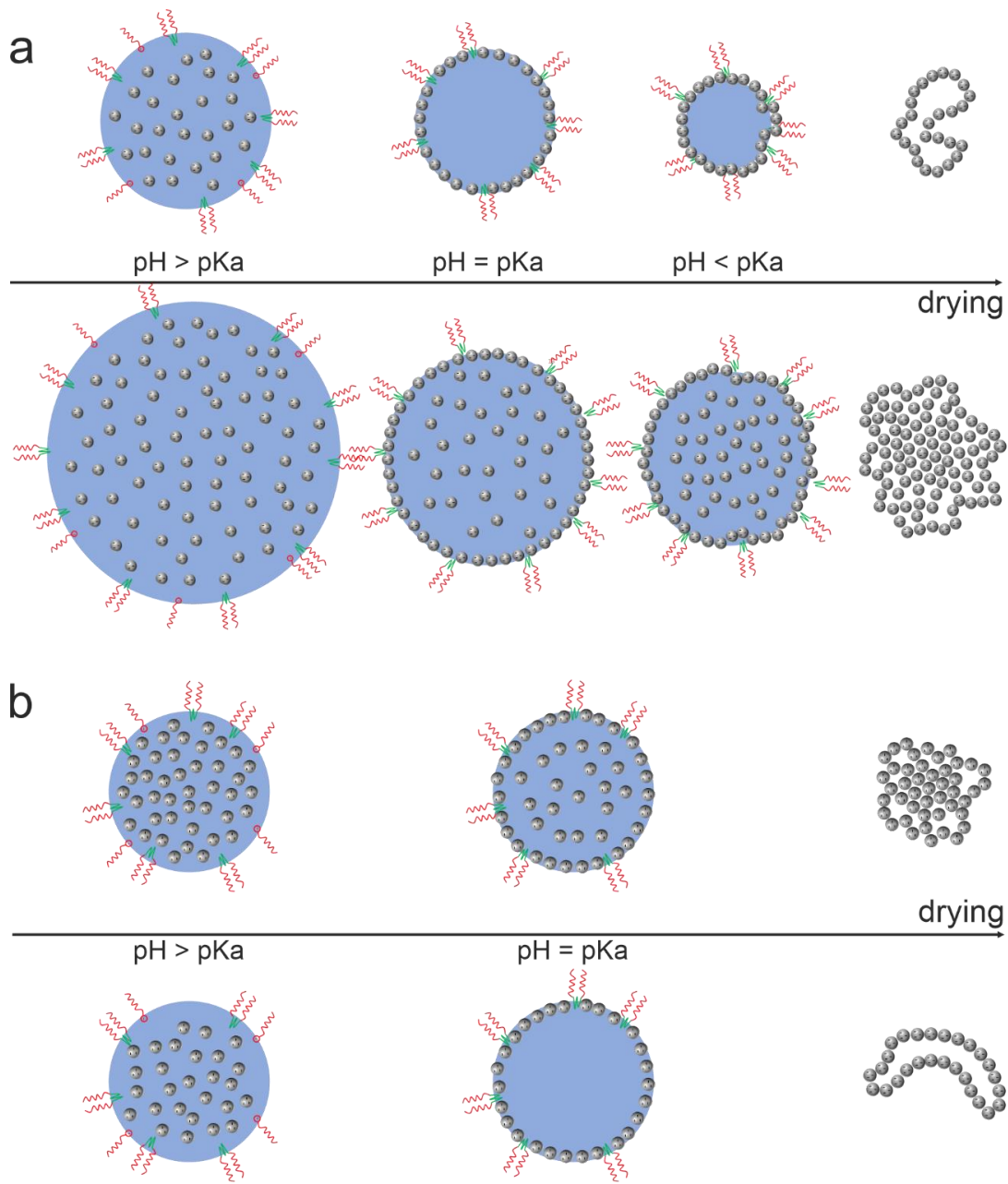

Figure S 7. Schematic representation of particle-laden aqueous droplets during drying. (a) Larger droplets have a higher proportion of particles left in the bulk after interfacial adsorption, resulting in less buckled structures. (b) For same size droplets, the particle concentration dictates the final morphology: A higher particle concentration leads to less buckled structures, since more particles are left in bulk after interfacial adsorption triggered by protonation of the surfactant.